\documentclass[aps,prx,reprint,floatfix,twocolumn,superscriptaddress]{revtex4-2}
\usepackage{graphicx,amsmath,bbm,amssymb}
\usepackage{appendix}
\usepackage{pgfplots}

\usepackage{braket}

\RequirePackage{color}
\definecolor{dkgreen}{rgb}{0.2,0.7,0.4}
\definecolor{dkblue}{rgb}{0.2,0.2,0.7}
\definecolor{dkred}{rgb}{0.8,0,0}
\definecolor{dkpurple}{rgb}{0.6,0.2,0.6}

\usepackage[english]{babel}
\begin{document}

\title{Photonic Analog Quantum Simulation of (1+1)-Dimensional $U(1)$ Lattice Gauge Theory with Dynamical Matter}

\author{Nathan R. Gonzalez}
\thanks{These authors contributed equally.}
\affiliation{Department of Physics and Astronomy, University of California, Davis, 1 Shields Avenue, Davis, CA, USA}
\affiliation{Department of Electrical and Computer Engineering, University of California, Davis, 1 Shields Avenue, Davis, CA, USA}
\author{Thea Budde}
\thanks{These authors contributed equally.}
\affiliation{Institut für Theoretische Physik, ETH Zürich, Wolfgang-Pauli-Str. 27, 8093 Zürich, Switzerland}
\author{Klemen Kersic}
\thanks{These authors contributed equally.}
\affiliation{Institut für Theoretische Physik, ETH Zürich, Wolfgang-Pauli-Str. 27, 8093 Zürich, Switzerland}
\author{Zia Steele}
\affiliation{Department of Physics and Astronomy, University of California, Davis, 1 Shields Avenue, Davis, CA, USA}
\affiliation{Department of Electrical and Computer Engineering, University of California, Davis, 1 Shields Avenue, Davis, CA, USA}
\author{Alex H. Rubin}
\affiliation{Department of Physics and Astronomy, University of California, Davis, 1 Shields Avenue, Davis, CA, USA}
\affiliation{Department of Electrical and Computer Engineering, University of California, Davis, 1 Shields Avenue, Davis, CA, USA}
\author{Joao C. Pinto Barros}
\affiliation{Institut für Theoretische Physik, ETH Zürich, Wolfgang-Pauli-Str. 27, 8093 Zürich, Switzerland}
\author{Marina Radulaski}
\affiliation{Department of Electrical and Computer Engineering, University of California, Davis, 1 Shields Avenue, Davis, CA, USA}
\author{Marina Krstic Marinkovic}
\affiliation{Institut für Theoretische Physik, ETH Zürich, Wolfgang-Pauli-Str. 27, 8093 Zürich, Switzerland}
\email{marinama@ethz.ch}

\begin{abstract}

    We propose a photonic scheme for analog quantum simulation of a $U(1)$ Lattice Gauge Theory (LGT) with dynamical matter based on the Jaynes-Cummings-Hubbard (JCH) model.
    Here, an array of interacting cavities in the strong-coupling regime of cavity Quantum Electrodynamics is mapped onto the alternating matter and gauge-field sites of the spin-1/2 Quantum Link Model. 
    In contrast to other analog LGT quantum simulation methods, our approach implements the desired gauge-invariant dynamics through the hopping of polaritonic excitations among the array sites. The hopping is mapped to the gauge theory via precise tuning of polaritonic resonances in individual cavities. 
    Using exact diagonalization, we show that the real-time evolution of the JCH model accurately replicates that of a Quantum Link Model.
    Finally, we discuss feasible routes to the beyond-classical simulation capability with scalable implementations in photonic and superconducting systems. 
    This provides a novel route towards understanding the real-time dynamics of lattice gauge theories with matter in higher dimensions.
\end{abstract}

\maketitle

\section{Introduction}
Quantum simulation offers a promising route to exploring strongly interacting quantum many-body systems that remain inaccessible to the existing classical approaches. This includes overcoming issues that arise either because Monte Carlo methods suffer from sign problems~\cite{Troyer2004,Alexandru2020}  or because entanglement growth makes tensor network representations prohibitively costly~\cite{Vidal:2003pmm, Schollwock:2005zz,Verstraete:2006mdr}. 
Gauge theories provide important examples of strongly interacting quantum systems in which these limitations arise directly in nonperturbative calculations. They underlie the Standard Model of particle physics~\cite{Gattringer2010,Bietenholz2024} and also arise as effective low-energy descriptions in condensed matter systems~\cite{Fradkin2013,Zhou_Kanoda_Ng_2017}.  Classical simulation approaches to lattice gauge theory (LGT) have been instrumental in precision Quantum Chromodynamics (QCD) calculations that serve as input for stringent tests of the Standard Model~\cite{FLAG24,Aliberti:2025beg}. However, many dynamical regimes of interest remain beyond the reach of existing classical approaches, including 
real-time dynamics~\cite{Halimeh:2025vvp} and strongly interacting matter at finite density~\cite{Nagata:2021ugx}. Further examples of classically intractable strongly interacting quantum many-body systems include systems other than QCD, such as frustrated quantum systems~\cite{Frustrated} and models commonly used for fault-tolerant quantum computing~\cite{Toric}. 

Quantum simulations of LGTs are underway on a variety of platforms~\cite{Halimeh:2025vvp,de_observation_2024,de_observation_2024,cobos_real-time_2025,gonzalez-cuadra_observation_2025,wang_observation_2025, Busnaina:2025gbk}. 
Most of the focus is on Abelian gauge groups, $\mathbb{Z}_N$ \cite{gorg_realization_2019,schweizer_floquet_2019,satzinger_realizing_2021,wang_observation_2022,cochran_visualizing_2025,luo_quantum_2025,mildenberger_confinement_2025,mueller_quantum_2025,Saner:2025nrq}
or $U(1)$ \cite{Marcos:2014lda, martinez_real-time_2016,dai_four-body_2017,klco_quantum-classical_2018,kokail_self-verifying_2019,yang_observation_2020,de_jong_quantum_2022,nguyen_digital_2022,zhou_thermalization_2022,farrell_quantum_2024,farrell_scalable_2024,zhu_probing_2024,Zhao:2025gxy,meth_simulating_2025,zhang_observation_2025, Alcaine-Cuervo:2026fgw}, with emerging progress towards $SU(N)$~\cite{Ciavarella:2024fzw, Balaji:2025afl,Das:2025utp, Yao:2025cxs, Froland:2025bqf, Li:2025sgo, Chen:2026hnh, Modi:2026syn, Chen:2026tvd,Webb-Mack:2026bkg}. 
Among the various platforms for quantum simulation, photonic cavity-, as well as superconducting circuit-Quantum Electrodynamics (QED), offer unique routes to integrate large-scale cavity-array-based quantum simulation architectures in solid-state platforms.

The standard formulations of gauge theories on the spatial lattice~\cite{wilson1974, kogut1975,Wilson_1977,Kogut_1978,Kogut_1979} entail infinite-dimensional Hilbert spaces.
This is a challenge because most quantum simulation platforms offer limited control over large-dimensional Hilbert spaces. 
A prominent approach to LGTs that requires only finite-dimensional Hilbert spaces is the Quantum Link Model (QLM) formulation \cite{Horn1981,orland1990,Chandrasekharan1997,Brower1997}, which will be used in this work. QLMs provide a framework in which each space point has a finite-dimensional quantum degree of freedom, while gauge symmetry is exactly preserved. QLMs offer a consistent formulation of the gauge theories in their own right, yet they
can be shown to reproduce the physics of the standard formulation of LGTs in the limit of large local Hilbert spaces \cite{Kasper2016,Zache2021}. 

\begin{figure*}
    \centering
    \includegraphics[width=0.8\linewidth]{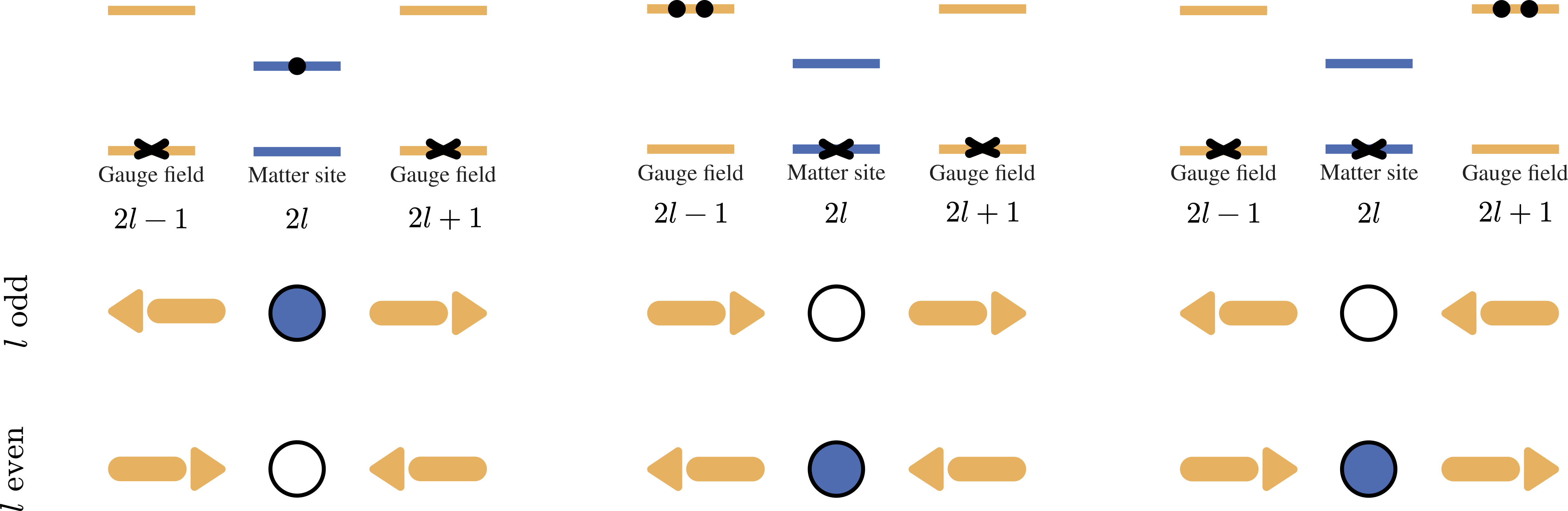}
    \caption{The $U(1)$ gauge theory can be mapped exactly onto a bosonic chain with $2L-1$ sites. The bosonic occupations are uniquely mapped to a gauge-theory configuration depending on the site's parity. The left/right pointing yellow arrows represent the polarization of the gauge field $\langle S^z _l\rangle = \pm 1/2$. A filled/empty circle indicates an occupied/empty fermionic site.  An occupied bosonic matter site corresponds to the presence of a charge in the QLM, while the location of doubly occupied gauge field sites indicates the polarization of the gauge field.
    }
    \label{fig:mapping}
\end{figure*}

In this work, we propose an experimentally feasible scheme for photonic analog quantum simulation of a ($1+1$)-dimensional $U(1)$ gauge theory, based on coupled cavity arrays modeled by the Jaynes-Cummings-Hubbard (JCH) Hamiltonian. By exploiting the anharmonic polaritonic spectrum of the Jaynes-Cummings model and engineering specific resonant conditions across a coupled array of cavities, we demonstrate that a $U(1)$-symmetric lattice gauge theory can emerge from the dynamics of the proposed system. 

The spin-1/2 quantum link model, which is a discretized formulation of the well-known Schwinger model \cite{Schwinger_1962} in the large coupling limit, is a $U(1)$ symmetric $(1+1)$-dimensional representation of Quantum Electrodynamics. In the QLM formulation, the Hamiltonian takes the form
\begin{equation}
    \hat{H} = t\sum_{l=0}^{L-2} \left(\psi^{\dagger}_lS^+_l\psi_{l+1}+h.c.\right)+m \sum_{l=0}^{L-1} (-1)^l\psi^\dagger_l\psi_l .
    \label{eq:H_qlm}
\end{equation}
The fermionic degrees of freedom $\psi_l$ live on the sites $l$ of a one-dimensional lattice and represent the matter degrees of freedom obeying the standard anti-commutation relation $\left\{\psi^\dagger_l,\psi_{l^\prime}\right\}=\delta_{ll^\prime}$. At each link connecting sites $l$ and $l+1$, there are spin degrees of freedom that represent the gauge fields. The operators $S^{\pm}_l$ are standard raising/lowering spin-1/2 operators. The first term in \mbox{Eq. \eqref{eq:H_qlm}} is the correlated hopping between matter and gauge fields, while the second term is the mass term.

This model is a $U(1)$ gauge theory, since it commutes with the local $U(1)$ operator $e^{i \phi_n G_n}$ where $\phi_n \in [0, 2\pi)$ and 
\begin{equation}\label{eq:GaussLaw}
    G_l = q_l  - (S^z_l - S^z_{l-1}) ,
\end{equation}
where the charge is defined as $q_l = \psi_l^\dagger \psi_l -\frac{1 + (-1)^l}{2}$. The Hilbert space is constrained to the physical sector, defined as the states that obey $G_l \ket{\psi} = 0$. This resembles the familiar Gauss's law in electrodynamics, in which the divergence of the electric field, now represented by spins, equals the local charge density at each point.

We show that this model arises in the low-energy, low-occupation subspace (average filling $\rho \leq 1$) of the JCH model, envisioned for implementation in superconducting circuit QED and color-center cavity QED systems. We verify this with numerical simulations employing specific resonant conditions, which show that a coupled cavity array can be engineered to emulate the dynamics of the desired $U(1)$ QLM, with effective three-body correlated hopping terms naturally arising. We derive an effective Hamiltonian via a Schrieffer-Wolff transformation that maps onto a gauge theory to second order in perturbation theory, providing an all-photonic route to simulate LGTs in an experimentally accessible manner.

The remainder of the paper is organized as follows. In Sec. \ref{sec:proposal}, we introduce the JCH model and derive the mapping to the spin-1/2 QLM. In Sec. \ref{sec:simulation}, we present exact diagonalization results that show the gauge-invariant real-time evolution of the JCH model. In Sec. \ref{sec:generalizations}, we discuss extensions to higher-dimensional gauge theories and models with larger local Hilbert spaces. In Sec. \ref{sec:Experiment}, we describe feasible experimental implementations of the proposed cavity array in both photonic cavity QED systems and superconducting circuit QED. We conclude in Sec. \ref{sec:conclusion}.

\section{Proposal}\label{sec:proposal}

In this section, we propose a scalable and versatile quantum simulation of gauge theories using experimental platforms well described by the Jaynes-Cummings-Hubbard model in the strong-coupling regime. 

\subsection{Mapping the QLM to a bosonic gauge theory}

\begin{figure*}
    \centering
    \includegraphics[width=0.65\linewidth]{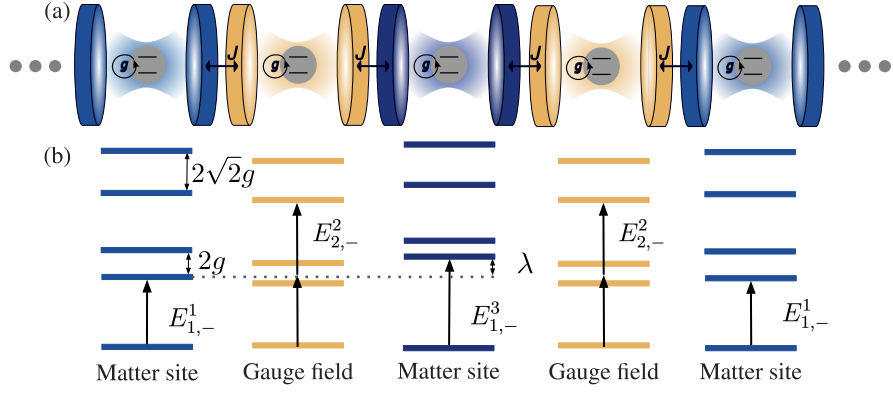}
    \caption{\textbf{Mapping between the JCH model and the spin-1/2 QLM.} (a) Schematic of a coupled cavity array in the strong coupling regime ($g\gg J$), consisting of alternating matter (blue) and gauge (orange) sites, each composed of a cavity coupled to a two-level emitter with a coupling rate $g$. Photon hopping between cavities occurs at a rate $J$. Colors indicate varying cavity frequency $\omega_e$. 
    (b) Polariton energy level structure for each cavity.
     The anharmonic splitting of the JC ladder is tuned to support the resonance conditions $E^{2i}_{2,-} = E^{2i-1}_{1,-} + E^{2i+1}_{1,-}$ to be satisfied simultaneously for all $i$ across the array.} 
    \label{fig:Mapping}
\end{figure*}

To map the physics of JCH systems onto the gauge theory Eq.~\eqref{eq:H_qlm}, we follow a mapping inspired by those used for simulations in systems governed by Bose-Hubbard models in Refs.~\cite{zhou_thermalization_2022, Majcen:2026sfm}. The QLM lattice composed of $L$ sites and $L-1$ links is converted to a lattice of $2L-1$ sites occupied by bosons. We associate the state of the gauge fields, which are represented by spins, with the absence or presence of a boson pair. This representation takes advantage of the equivalence of hard-core bosons, fermions, and spins in one dimension, which can be established by a Jordan-Wigner transformation. Explicitly, the map distinguishes between $l$ even or odd according to  $\hat{\psi}_l \to b_{2l}^\dagger$ and  $S^+_l \to (b_{2l+1}^\dagger)^2$ for even $l$ and $\psi_l \to b_{2l}$ and  $S^+_l \to (b_{2l+1})^2$ for odd $l$. This correspondence is illustrated in Fig.~ \ref{fig:mapping}. This results in the Hamiltonian
\begin{equation}
    \label{eq:Heff}
    H =  \frac{t}{\sqrt{2}}\sum_{l=0}^{L-1} \left( b_{2l} \left(b^\dagger_{2l+1}\right)^2 b_{2l+2} + h.c. \right) +  m \sum_{l=0}^{L-1} b^\dagger_{2l} b_{2l}. 
\end{equation}
To make this map consistent, we must ensure that the system is initialized in the subspace corresponding to a physical state in the QLM. Explicitly, this means that even sites of the bosonic chain are initialized with occupation 0 or 1, odd sites are initialized with occupation 0 or 2, no two neighboring sites are occupied, and there are a total of $L$ bosons. We note that the bosonic chain in Eq.~\eqref{eq:Heff} is also a (different) $U(1)$ gauge theory, since it commutes with the operators
\begin{equation}\label{eq:bosonicGaussLaw}
    \tilde{G}_{2l} = b^\dagger_{2l-1} b_{2l-1} + 2b^\dagger_{2l} b_{2l} + b^\dagger_{2l+1}b_{2l+1} - 2.
\end{equation}
In particular, the physical sector of the original QLM is mapped exactly to the sector $\tilde{G}_{2l} \ket{\psi} = 0$.
In this representation, an occupied even matter site represents the presence of a negative charge, an occupied odd site represents the presence of a positive charge, and an unoccupied site represents no charge at that site. This representation of the truncated Schwinger model has the advantage of explicit particle-number conservation in the interactions, while consisting of three rather than four-body interactions, as in Rishon formulations.

The Hamiltonian Eq.~\eqref{eq:Heff} can be implemented in a quantum simulator governed by the Jaynes-Cummings-Hubbard model, by engineering a resonance condition in the strong-coupling regime, as described below.

\subsection{Jaynes-Cummings-Hubbard Model}

The Jaynes-Cummings-Hubbard (JCH) model is described by the Hamiltonian
\begin{align}
    \label{eq:H_JCH}
    \hat{H}_{JCH} = &\sum_i \left(\omega^c_ia^{\dagger}_ia_i +\omega^e_i\sigma^+_i\sigma^-_i+g_i\left(a_i\sigma^+_i+h.c.\right) \right)\nonumber \\
    &+ \sum_{\langle i,j\rangle} J_{ij} \left(a^\dagger_i a_j + h.c.\right)
\end{align}
where for a given site $i$, $a_i^{(\dagger)}$ are the bosonic annihilation (creation) operators for the cavity mode, $\sigma^{+(-)}_i$ are spin raising (lowering) operators of the emitter, $g_i$ is the emitter-cavity coupling rate, $\omega^{c(e)}_i$ is the cavity (emitter) frequency, and $\langle i,j\rangle$ denotes pairs of nearest neighbor sites on a general lattice. Systems that are well described by this model include photonic cavity QED and superconducting circuit QED, both discussed in Sec. \ref{sec:Experiment}.

Neglecting the coupling $J_{ij}$ between cavities, the polaritonic eigenenergies of each site are $E^i_0 = 0$
and
\begin{equation}
    \label{eq:H_JC-energies}
    E^i_{n_i \geq 1,\pm} = \omega^c_i n_i + \frac{\Delta_i}{2} \pm \sqrt{\left(\frac{\Delta_i}{2}\right)^2 + g_i^2n_i},
\end{equation}
where $n_i = a^\dagger_i a_i + \sigma^+_i\sigma^-_i$ is the number of excitations at the site $i$ and $\Delta_i = \omega^e_i - \omega^c_i$ is the detuning between cavity and emitter. The $\pm$ branches define the upper and lower polariton manifolds. The level spacing between these manifolds is anharmonic, allowing fine-tuning of resonance conditions.

\begin{figure*}[htb!]
    \centering
    \includegraphics[width = \linewidth]{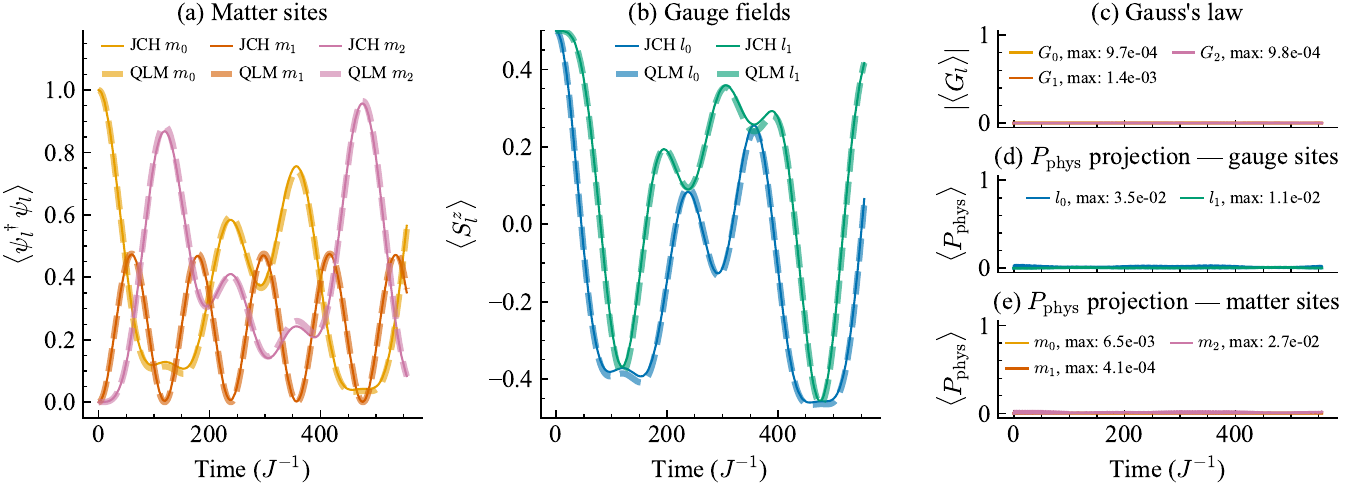}
    \caption{\textbf{Real-time dynamics for the JCH model on a chain of five coupled cavities compared to the spin-1/2 QLM.}
    (a) Matter site and (b) gauge field JCH occupation mapped to QLM observables (charge density and electric flux, respectively) as described by Eqs. \eqref{eq:fermion_map} and \eqref{eq:spin_map}.
    (c) Absolute value of the expectation value for the Gauss's law operator $|\langle G_l\rangle|$ for each matter site $l$ and the two adjacent gauge sites, which remains close to zero for all times, confirming confinement to the physical subspace by the gauge protection term $\lambda$. Gauss's law $G_l$ is defined in Eq. \eqref{eq:bosonicGaussLaw}. 
    Projection onto the physical state for (d) gauge fields and (e) matter sites as described by the set of equations in Eq. \eqref{eq:projectors}.
    Simulations are done for parameters $\omega_e/J = 500$, $\Delta_0/J=111.1$, $g/J = 333.3$, and $\lambda/J = 40$, which correspond to $t/J=0.02021$ and $m/J=0.0064389$ in the QLM.
    } 
    \label{fig:JCH_QLM_comp}
\end{figure*}

\subsection{Mapping the JCH model to the gauge theory}

To engineer the Hamiltonian Eq. \eqref{eq:Heff}, the parameters of the JCH Hamiltonian are tuned such that the resonance conditions 
\begin{equation}
    E_{2, -}^{2i} = E_{1, -}^{2i-1} + E_{1, -}^{2i+1}, \quad E_{1, -}^{2i-1} = E_{1, -}^{2i+1} + \lambda_{2i}
\end{equation}
are obeyed, where $|\lambda_{2i}| \gg J_{ij}$ are free parameters. The first condition ensures that the correlated two-polariton process, in which two excitations are absorbed by a gauge site where one comes from each neighboring matter site, is resonant. The second equation introduces an energy offset between neighboring matter sites, which prevents an effective gauge-violating interaction $(b_{2l}^\dagger b_{2l+2}+h.c.)$.

We set $\omega^e_i = \omega_e$, $g_i = g$, and $J_{ij} = J$ to be homogeneous across all cavities, while varying the cavity detuning $\Delta_i$, the most freely tunable parameter in experimental settings. To ensure a homogeneous value of $t$, we choose $\lambda_{2l} = (-1)^l \lambda_0$ to be staggered. To fulfill this resonance condition, three different cavity detunings $\Delta_i$ are required, which repeat in a periodic pattern, as depicted in Fig.~\ref{fig:Mapping}.
If the system is initialized within the physical subspace, the desired effective Hamiltonian Eq. $\eqref{eq:Heff}$ emerges to order $\mathcal{O}((J/\varepsilon)^2)$, where $\varepsilon$ is the smallest energy gap between a physical state and a non-physical state, in the limit $J \to 0$. The gap $\varepsilon$ is a function of all system parameters, but generally increases with $g$ and $\lambda$ if $\omega_e$ is large. Therefore, the strong coupling regime $g \gg J$, and large inhomogeneity $\lambda \gg J$ is required.

\section{Real-time dynamics} \label{sec:simulation}

To verify that the JCH system dynamics reproduce those of the spin-1/2 QLM, we perform exact diagonalization on both Hamiltonians and compare their real-time evolution directly. Agreement between the two models is established by comparing the dynamics of local observables (Figs. 3 a,b) and by verifying that the system remains in the physical subspace (Fig. 3 c,d,e).

\subsection{Comparing JCH and QLM dynamics}

Figure \ref{fig:JCH_QLM_comp} shows the real-time evolution comparison between the full JCH model and spin-1/2 QLM for parameters $\omega_e/J = 500$, $\Delta_0/J=111.1$, $g/J = 333.3$, and $\lambda/J = 40$. The cavity modes are restricted to hold at most $N_c = 3$ photons to reduce the simulation runtime.

To make comparisons, we map observables from one model to the other. On matter sites, this mapping is given by
\begin{equation}\label{eq:fermion_map}
    \langle \psi_l^\dagger \psi_l\rangle \equiv 
    \begin{cases}
        \langle 1 - n_{2l} \rangle & \text{for even $l$}, \\
        \langle n_{2l} \rangle & \text{for odd $l$}, \\
    \end{cases}
\end{equation}
and for gauge links,
\begin{equation}\label{eq:spin_map}
    \langle S^z_{l} \rangle \equiv
    \begin{cases}
        \langle \frac{1}{2}(n_{2l + 1} - 1) \rangle & \text{for even $l$}, \\
        \langle \frac{1}{2}(1 -n_{2l + 1}) \rangle & \text{for odd $l$}. \\
    \end{cases}
\end{equation}
Where in each case, the total occupation at each JCH site $l$ is given as $n_{l} = a_l^\dagger a_l + \sigma_l^+\sigma_l^-$, which counts both excitations in the cavity and emitter modes.

\begin{figure}
    \centering
    \includegraphics[width=\linewidth]{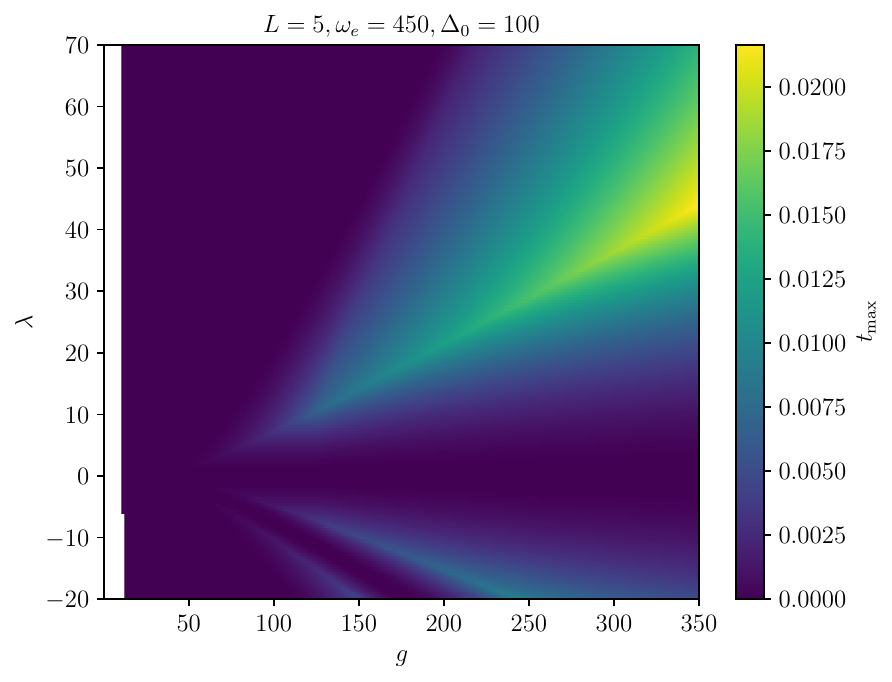}
    \caption{
    The possible strength of the effective interaction $t$ varies with the parameters. The color indicates the maximum value of $t$ that can be engineered while keeping $J$ perturbatively small. This ensures gauge invariance at intermediate timescales. White indicates that there is no valid solution to the resonance condition for these parameters.
    }
    \label{fig:operating_regime}
\end{figure}

The system studied in Fig. 3 is prepared in the initial state, which in the JCH model is $\ket{n_0, n_1, n_2, n_3, n_4} = \ket{0, 2^-, 0, 0, 1^-}$ and corresponds to $\ket{\psi_0^\dagger\psi_0,S^z_0,\psi_1^\dagger\psi_1,S^z_1,\psi_2^\dagger\psi_2} = \ket{0, \frac{1}{2}, 0, -\frac{1}{2}, 1}$ in the QLM. The observables measured in the evolution of the JCH model closely track those in the QLM for both matter (Fig. 3a) and gauge (Fig. 3b) field occupations. To quantify gauge invariance and how well confined to the physical subspace the system is throughout the evolution, we track the expectation value of Gauss's law of Eq.~\eqref{eq:bosonicGaussLaw} in Fig. 3(c). As expected, it remains negligible at all times for suitably adjusted parameters. Additionally, we verify that the occupation of polariton states that are not associated with states in the original gauge theory remains negligible by measuring the projection onto the physical states on each site, given by
\begin{equation}\label{eq:projectors}
    P^{\mathrm{phys}}_l = 
    \begin{cases}
        1-\ket{0}\bra{0} - \ket{1^-}\bra{1^-} & \text{for matter sites}, \\
        1-\ket{0}\bra{0} - \ket{2^-}\bra{2^-} & \text{for links}.
    \end{cases}
\end{equation}
These projections are shown in Fig.~3(d) and (e), which show that the system remains in the physical subspace to a good approximation throughout the evolution.

These results show that the JCH model with appropriate resonance engineering is a promising platform for analog quantum simulation of the real-time dynamics of the spin-1/2 QLM.

\subsection{Parameter regime}

The parameters of the effective Hamiltonian Eq.~\eqref{eq:Heff} depend on the JCH parameters. They can be determined through a second-order Schrieffer-Wolff transformation.
Fig. ~\ref{fig:operating_regime} shows the maximal effective hopping rate $t$ in Eq.~\eqref{eq:Heff} that can be achieved while retaining gauge invariance at intermediate timescales for a range of parameters. To ensure gauge invariant dynamics at intermediate timescales, we choose a maximum value of $J = \varepsilon/15$, where $\varepsilon$ is the minimum gap between a physical and a non-physical state.

Certain combinations of parameters optimize the effective hopping $t$ and lead to faster gauge-invariant dynamics. In an experimental implementation, dissipation should be slower than $t$.

\section{Generalizations}\label{sec:generalizations}

\subsection{(2+1)-dimensional lattice gauge theory}

The proposal can be extended to arbitrary lattice geometries, such as periodic boundaries and higher-dimensional lattices. Similar proposals have been made in Refs.~\cite{Osborne:2022jxq, Majcen:2026sfm} On a lattice with bosonic degrees of freedom on the vertices $b_i$ and on the links connecting two vertices $b_{i,j}$, the model
\begin{equation}
    H =  \frac{t}{\sqrt{2}}\sum_{\langle i,j\rangle}  \left( b_{i} \left(b^+_{i,j}\right)^2 b_{j} + h.c. \right) +  m\sum_{i} b_{i}^\dagger b_i
\end{equation}
is a gauge theory.
Here, the first sum goes over all links and the second over all sites. This model has the $U(1)$ gauge symmetry
\begin{equation}
    G_i = 2 b_i^\dagger b_i + \sum_{\langle i,j \rangle} b_{i,j}^\dagger b_{i,j}
\end{equation}
where the sum goes over all links connected to site $i$. This model can be mapped to one that closely resembles the $(2+1)$-dimensional $S=1/2$ QLM, without pure-gauge terms, in analogy to the $(1+1)$-dimensional case. However, since the Jordan-Wigner transformation does not locally map fermions to hard-core bosons in higher dimensions, the model has hard-core bosonic matter rather than fermionic matter.

We ensure $E_{2, -}^{i,j} = E_{1, -}^{i} + E_{1, -}^{j}$ and $E_{1, -}^{i} - E_{1, -}^{j} = \lambda_{j,k}$ for all connected triplets of sites $i$,$k$ and links $j$. The anisotropy $\lambda_{j,k} \gg J_{i,j}$ can be inhomogeneous but must be large enough to suppress photons from tunneling between matter sites. For square lattices, choosing a staggered lambda will once again lead to a homogeneous $t$.

\begin{figure*}[htb!]
    \centering
    \includegraphics[width=0.7\linewidth]{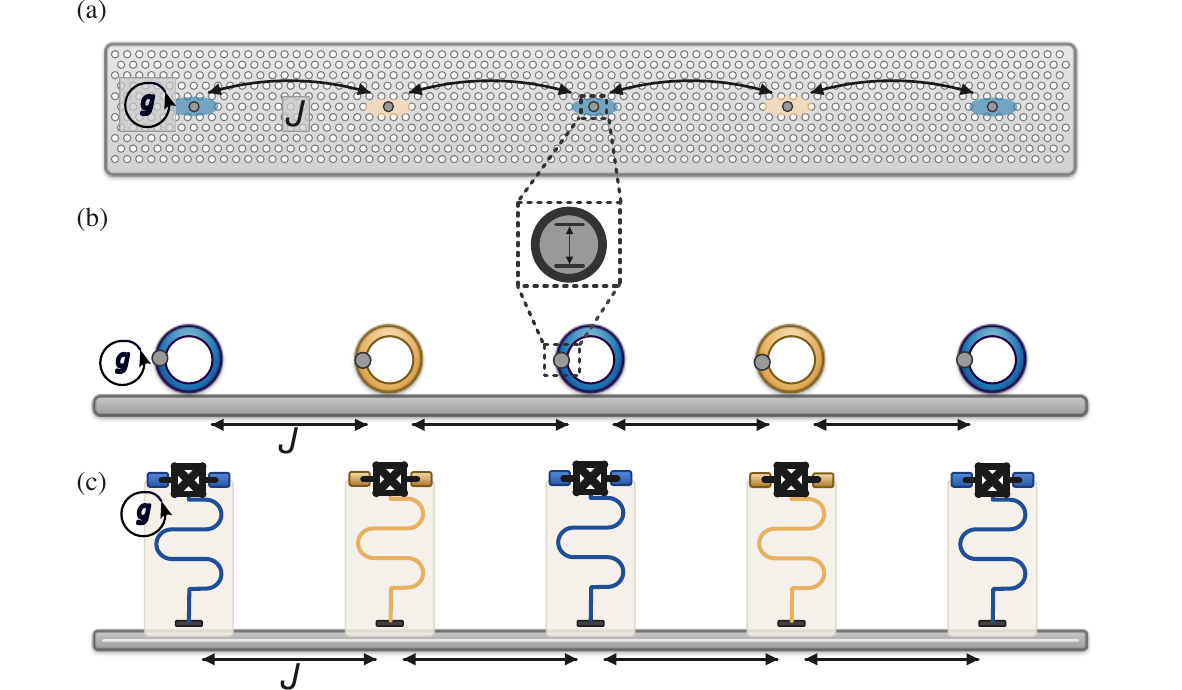}
    \caption{\textbf{Proposed experimental implementations of the analog quantum simulation.} Blue is used to represent matter sites and yellow to represent gauge fields. (a) Coupled photonic crystal cavity array, with each cavity coupled to a quantum emitter. (b) Optical whispering gallery mode resonators coupled to quantum emitters. The resonators are evanescently coupled to a waveguide to allow for hopping. In panels (a) and (b), the quantum emitters to which the photonic cavities are coupled could be color centers or quantum dots. (c) Superconducting qubits coupled to meandered coplanar waveguide resonators to form the local emitter-cavity JC system. The resonators are capacitively coupled to a transmission line to generate intercavity hopping.
    }
    \label{fig:CCA}
\end{figure*}

\subsection{Gauge theories with large local Hilbert spaces}

By placing harmonic cavities (without an emitter) to represent the gauge fields, one can realize gauge theories with high-dimensional local Hilbert spaces. By ensuring these cavities frequencies are resonant $\omega_{jk} = E_{1, -}^{j} + E_{1, -}^{k}$, while suppressing next to nearest neighbor hoppings with $E_{1, -}^{j} = E_{1, -}^{k} - \lambda_{j,k}$, and $\omega_{jk} = \omega_{jl} + \tilde{\lambda}_{jkl}$, where all $\lambda, \tilde{\lambda} \gg J$,  we ensure that the dynamics are gauge invariant as long as the link sites are initialized in an even occupation $n$. Note that the bosonic operators obtain different prefactors from spins and that the pure gauge term is missing in this simulation. Therefore, this is not the Schwinger model, but another lattice gauge theory that does not necessarily have a continuum limit. If ultrafast nonlinear control is available on the platform, the bosonic mode can be modulated in a Trotter-step fashion to realize the pure gauge term and recover the same physics in the limit of large occupation \cite{Davoudi:2021ney}.

\section{Experimental realizations}\label{sec:Experiment}
One of the major advantages of the proposed model is its implementation across a variety of technologies capable of performing JCH interactions. In the following, we discuss implementations in photonic cavity QED and superconducting circuit QED platforms.

\subsection{Photonic cavity QED realization}
Photonic systems are a natural choice for the implementation of the proposed analog quantum simulator of LGTs. Solid-state systems provide the scalability needed to achieve the computational advantage over classical computing resources, while the spatial light modulators \cite{lin2025optical, panuski2022full} and 2D single-photon imaging cameras support simultaneous excitation and readout of separate array sites. Quantum emitters can be selected between 1) individual quantum dots, which provide strong emitter-cavity interaction rates \cite{hennessy2007quantum, majumdar2012probing}, but can be challenging to keep homogeneous along the array, 2) single color centers, which have recently been shown to strongly couple to a photonic cavity \cite{lukin2025mesoscopic}, or 3) an ensemble of color centers, where the simulator would be described by the Tavis-Cummings-Hubbard model \cite{patton2024polariton, mondal2024emergence}, requiring an adjustment in resonant energies compared to the JCH model discussed here. Coupled cavity arrays can be formed by photonic crystal cavities \cite{majumdar2012cavity, Komza_2025} or whispering gallery mode resonators \cite{saxena2023realizing} coupled by direct proximity or via a common waveguide, as illustrated in Fig. \ref{fig:CCA}a-b. Implementations in wafer-scale substrates that host quantum emitters, such as silicon, silicon carbide and III-Vs or via hybrid integration of defects in 2D materials on taped-out samples, could support up to $\sim 10^4$ cavities. Finally, the polariton resonance engineering can be done by passive tuning of the individual nodes, e.g., by the permanent laser oxidation of material, which irreversibly tunes cavity resonance \cite{Chen_2011, Piggott_Vuckovic_2014}.

\subsection{Superconducting circuit QED realization}
Superconducting circuit QED systems are another platform well-suited to the implementation of the proposed analog quantum simulator of LGTs, albeit in the microwave domain.
In this context, the simulator would consist of a linear chain of coupled microwave resonators, each coupled to a transmon qubit.
Scalability to intermediate system sizes is well-established in this platform, with current processors demonstrating hundreds of high-coherence qubits.
Benefits include the independent dispersive readout of individual qubits via dedicated feedlines \cite{sc_review}, which naturally supports simultaneous state measurement across the array.
Coupled cavity arrays can be formed using coplanar waveguide (CPW) resonators or lumped-element LC circuits, coupled capacitively in direct proximity to form nearest-neighbor tight-binding chains \cite{transmon_chain}, as illustrated in Fig. \ref{fig:CCA}c.
Such arrays of Josephson-junction resonators interacting with transmons have recently been leveraged to successfully demonstrate the emergence and control of atom-photon bound states \cite{superconducting_bound}.
Superconducting circuits operate at microwave frequencies with correspondingly larger spatial footprints than nanophotonic systems, potentially limiting ultimate integration density and thus the maximum number of nodes in the simulator.
These circuits also require deep cryogenic temperatures in the milliKelvin range, in contrast to optical quantum emitters which can often operate at higher cryogenic temperatures in the 1-10 K range. 
However, this platform offers the distinct advantage of extensive in-situ tunability.
Specifically, frequency-tunable transmons and tunable coupling schemes enable precise polariton resonance engineering to be performed dynamically \cite{sc_tunable_1, sc_tunable_2}.
This robust active control, in contrast to the permanent passive tuning methods required in photonic implementations, makes it straightforward to reconfigure the simulator's interaction parameters and effectively compensate for fabrication variances without fabricating a new device.
State-of-the-art dilution refrigerator systems and multi-layer wafer-scale packaging can presently support arrays scaling to several hundred nodes \cite{sc_qubit_number}.

\section{Conclusions}\label{sec:conclusion}
We have proposed and analyzed a photonic analog quantum simulation of a $(1+1)$-dimensional, $U(1)$ lattice gauge theory based on the Jaynes-Cummings-Hubbard model. Gauge-invariant dynamics emerge from resonant polaritonic hopping in a coupled cavity array, which can be mapped to the correlated matter-gauge hopping of the spin-1/2 Quantum Link Model. We establish that the JCH model effectively describes gauge-invariant dynamics to second order in degenerate perturbation theory. We also verify the mapping to the QLM numerically via exact diagonalization. We have shown that gauge invariance, as quantified by Gauss's law, is well preserved throughout real-time evolution. 

This approach is compatible with any system that can realize the JCH Hamiltonian in the strong coupling regime, but is particularly applicable to both superconducting circuit QED and photonic cavity QED. These platforms are particularly appealing for their potential to enable large-scale quantum simulations that push far beyond classical limits. For example, photonic crystal cavity arrays could support up to ~$10^4$ cavities, while superconducting circuits benefit from extensive in situ tunability and could support arrays of several hundred cavity-emitter nodes. Superconducting circuits offer clear advantages in system parameters with current technologies, and photonic systems offer potential for long-term scalability. Both platforms offer realistic routes to intermediate scale beyond classical analog quantum simulation.

Beyond the closed-system treatment given here, an important question for experimental implementation is the effect of cavity losses and emitter decay, which are absent from the closed-system treatment presented here. In both platforms, the strong coupling condition $g\gg \kappa_{\text{loss}},\gamma$ is required, both for the described polariton dynamics and to maintain coherent dynamics on timescales comparable to $J^{-1}$. State-of-the-art photonic platforms may be able to achieve $g/\kappa_{\text{loss}} \approx1-10$ \cite{lukin2025mesoscopic}, while superconducting circuits approach $g/\kappa_{\text{loss}}\approx 10-1,000+$ \cite{sc_review}.
The proposed platforms would provide a route for studying anomalous thermalization and other out-of-equilibrium phenomena, such as hadronization, in large volumes. 

We have further outlined potential extensions to this approach by discussing application to a $(2+1)$-dimensional lattice gauge theory and a model with a larger local Hilbert space. Together, these results establish coupled cavity QED as a promising platform for quantum simulation and provide a clear path toward intermediate and large-scale simulation of physics well beyond the capabilities of classical computation.

\section{Acknowledgments}

MR, NRG, ZS and AHR acknowledge support from the National Science Foundation CAREER award (No. 2047564). MR acknowledges support by the ETH Quantum Center Faculty Fellowship. MKM thanks the Kavli Institute for Theoretical Physics (KITP) for hospitality and support during the program “What is Particle Theory?” The KITP is supported in part by the National Science Foundation under Grant No. PHY-2309135. MKM is grateful for the hospitality of Perimeter Institute, where part of this work was carried out. Research at Perimeter Institute is supported in part by the Government of Canada through the Department of Innovation, Science and Economic Development and by the Province of Ontario through the Ministry of Colleges and Universities. This research was also supported in part by the Simons Foundation through the Simons Foundation Emmy Noether Fellows Program at Perimeter Institute.

\section{Author Contributions}

MR and MKM conceptualized the research problem and supervised research, JPB conceptualized theoretical aspects, KK, TB and NRG performed theoretical derivations, NRG, ZS, AHR, TB and KK performed numerical simulations, NRG, TB, JPB wrote the first draft of the paper, all authors contributed to editing and proofreading of the paper.

\bibliography{refs}

\end{document}